\documentclass{article}

\usepackage{microtype}
\usepackage{graphicx}
\usepackage{float}
\usepackage{cuted}
\usepackage{capt-of}
\usepackage{booktabs}
\usepackage{array}
\usepackage{tabularx}
\usepackage{arydshln}
\usepackage{xcolor}
\usepackage{amsmath}
\usepackage{amssymb}
\usepackage{mathtools}
\usepackage{enumitem}
\usepackage{xspace}
\usepackage{tikz}
\usetikzlibrary{arrows.meta,positioning,fit,shapes.geometric,decorations.pathreplacing}
\usepackage{pgfplots}
\pgfplotsset{compat=1.18}
\usepackage[hyperfootnotes=false]{hyperref}

\usepackage[accepted]{mlsys2025}
\makeatletter
\renewcommand{\Notice@String}{}
\renewcommand{\mlsys@appearing}{}
\makeatother

\newcommand{\nvidialogoaboverule}{%
  \nointerlineskip\vbox to 0pt{\vss
    \hbox{\includegraphics[width=0.8in]{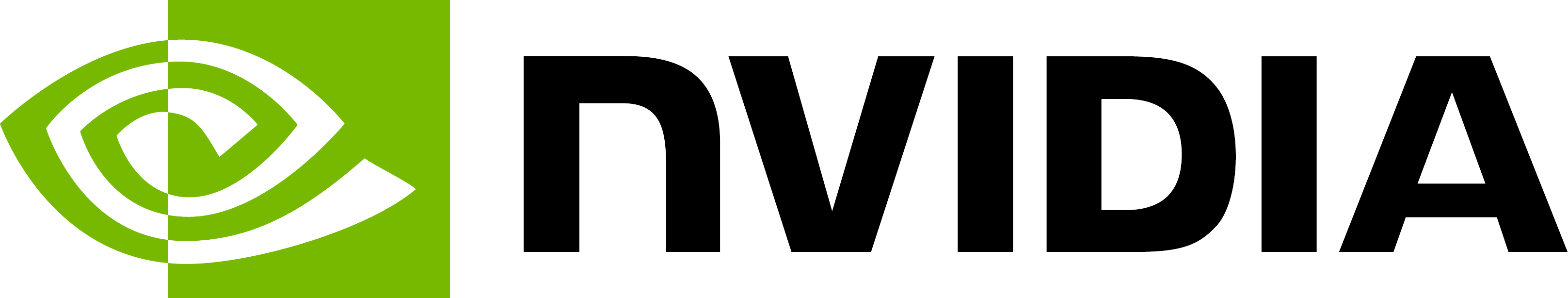}}%
    \vskip 2pt}}
\let\originaltoptitlebar\toptitlebar
\renewcommand{\toptitlebar}{\nvidialogoaboverule\originaltoptitlebar}
\let\originalheadrule\headrule
\renewcommand{\headrule}{\nvidialogoaboverule\originalheadrule}

\hypersetup{
  pdftitle={Serving a Revisable World: Versioned Execution for Interruptible Agents},
  pdfauthor={Yanxin Zhang, Rahul Sharma, Nitin Vegesna, Zheyu Fu, Chang Liu, Trivikram Krishnamurthy},
  pdfsubject={Versioned execution for interruptible agent serving},
  pdfkeywords={LLM agents, revisable execution, inference serving, KV cache}
}

\newlength{\mlsysTextFloatSep}
\newlength{\mlsysDblTextFloatSep}
\mlsystitlerunning{\smash{Serving a Revisable World}}

\newcommand{\retire}{\textsc{Retire}\xspace}

\begin{document}

\twocolumn[
\mlsystitle{\texorpdfstring{Serving a Revisable World:\\
Versioned Execution for Interruptible Agents}{Serving a Revisable World: Versioned Execution for Interruptible Agents}}

\begin{mlsysauthorlist}
\mlsysauthor{Yanxin Zhang}{nvidia}\hspace{1em}
\mlsysauthor{Rahul Sharma}{nvidia}\hspace{1em}
\mlsysauthor{Nitin Vegesna}{nvidia}\par\vspace{3pt}
\mlsysauthor{Zheyu Fu}{nvidia}\hspace{1em}
\mlsysauthor{Chang Liu}{nvidia}\hspace{1em}
\mlsysauthor{Trivikram Krishnamurthy}{nvidia}
\par\vspace{4pt}
{\normalfont\textsuperscript{1}NVIDIA Corporation, Santa Clara, CA, USA}
\end{mlsysauthorlist}

\mlsysaffiliation{nvidia}{NVIDIA Corporation, Santa Clara, CA, USA}
\mlsyscorrespondingauthor{Yanxin Zhang}{\href{mailto:yanxinz@nvidia.com}{yanxinz@nvidia.com}}

\mlsyskeywords{LLM agents, revisable execution, versioned serving,
supersession, inference serving, KV cache, continuous batching}

\vskip 0.1in
\begin{abstract}
LLM agents revise running tasks when users change instructions, tools fail,
or new information changes a plan.  Today's servers express a revision as
aborting old requests and submitting replacements.  Yet the old execution's
buffered output and outstanding work must stop affecting the application,
while completed KV state may still be useful to its replacement.  Handling
these obligations separately can leave obsolete effects publishable and force
the successor to rebuild valid state.

We present \retire, a serving control-plane redesign around \emph{versioned
execution}.  Requests own scheduling and memory resources; execution versions
own \emph{authority}, the permission to publish output or install state for
the current execution.  \retire first revokes obsolete work, then bounds its
remaining execution and certifies the completed prefix its successor can
inherit.  The successor runs from that state while isolated old resources
are reclaimed asynchronously.  This unifies fast invalidation and selective
preservation in one version transition.

We implement \retire in vLLM across output publication, GPU execution, KV
handoff, tiered recovery, and distributed and multi-tenant serving.  Correctness
experiments verify current-version output and valid state inheritance across
these paths.  Combining invalidation with inheritance reduces
revision-to-successor time-to-first-token by a median 17.1\% in controlled
paired experiments.  A replay of recorded coding-agent interruption arrivals
emits no obsolete output and keeps every final version progressing through
repeated revisions.  \retire turns abort-and-restart into a coordinated handoff
that stops obsolete work quickly and preserves useful work for its successor.
\end{abstract}

]

\begingroup
\renewcommand{\thefootnote}{}
\footnotetext{\raggedright Corresponding author:
\mbox{Yanxin Zhang}
\mbox{\textless\href{mailto:yanxinz@nvidia.com}{yanxinz@nvidia.com}\textgreater}.}
\endgroup

\section{Introduction}
\label{sec:introduction}

Consider a coding agent implementing an API change.  While the model is
processing repository context, the user asks it to keep the existing API and
change only the implementation.  The agent runtime replaces the plan.  Output
buffered for the abandoned plan must no longer reach the user, and its remaining
model work has become obsolete.  Yet the repository context shared by both
plans still has value: its completed KV prefix can accelerate the revised
execution.  One interruption therefore asks the server to stop obsolete work
and preserve useful work at the same time.

Such revisions are a normal part of agent execution.  A tool failure can
invalidate a plan, and a fresh observation can replace the assumptions behind
ongoing reasoning.  The agent runtime chooses the response and constructs the
revised context.  Here, \emph{authority} means permission for output or state
updates to affect the application or the successor.  Once the runtime decides
to replace live work, the server must carry out one decision:

\begingroup
\setlength{\leftskip}{1.8em}
\setlength{\rightskip}{1.8em}
\setlength{\parskip}{2pt}
\interlinepenalty=10000
\noindent\itshape
Revoke the old execution's authority and let its successor take over
from the state that remains valid.\par
\endgroup

\begin{figure}[t]
  \centering
  \resizebox{0.98\columnwidth}{!}{%
  \begin{tikzpicture}[
      font=\scriptsize,
      cause/.style={draw,rounded corners=5pt,fill=gray!7,align=center,
                    minimum height=6.5mm,minimum width=12mm,inner xsep=1.2mm},
      decision/.style={draw,rounded corners=3pt,fill=blue!10,align=center,
                       minimum height=9mm,minimum width=33mm},
      event/.style={draw,rounded corners=7pt,fill=blue!4,draw=blue!55!black,
                    align=center,minimum height=7mm,minimum width=43mm},
      authority/.style={draw,rounded corners=3pt,fill=orange!10,
                        draw=orange!65!black,align=center,
                        minimum height=10mm,minimum width=54mm},
      component/.style={draw,rounded corners=5pt,fill=green!7,
                        draw=green!45!black,align=center,
                        minimum height=6.5mm,minimum width=10mm,
                        inner xsep=1mm},
      arr/.style={-{Latex[length=1.7mm]},thick},
      merge/.style={draw=gray!55,semithick},
      fan/.style={-{Latex[length=1.3mm]},draw=orange!65!black,semithick}
    ]
    \node[cause] (u) at (0.55,5.40) {User};
    \node[cause] (t) at (1.85,5.40) {Tool};
    \node[cause] (v) at (3.15,5.40) {State};
    \node[cause] (p) at (4.45,5.40) {Planner};
    \draw[merge] (u.south)--(0.55,4.95);
    \draw[merge] (t.south)--(1.85,4.95);
    \draw[merge] (v.south)--(3.15,4.95);
    \draw[merge] (p.south)--(4.45,4.95);
    \draw[merge] (0.55,4.95)--(4.45,4.95);

    \node[decision] (d) at (2.50,4.25)
      {\textbf{Agent runtime}\\chooses $X'$};
    \draw[arr] (2.50,4.95)--(d.north);
    \node[event] (s) at (2.50,3.30)
      {$\mathsf{SUPERSEDE}(s,e,e{+}1,X',C)$};
    \draw[arr] (d)--(s);

    \draw[densely dashed,thick,gray!55] (-0.20,2.72)--(5.65,2.72);
    \node[fill=white,inner xsep=4pt,inner ysep=0.4pt,
          font=\scriptsize\bfseries,text=gray!65]
      at (4.35,2.72) {agent / serving};

    \node[authority] (a) at (2.50,1.95)
      {\textbf{\retire: authority switch}\\[-0.5mm]
       \textcolor{red!70!black}{$e$: revoked}
       $\;\longrightarrow\;$
       \textcolor{green!40!black}{$e{+}1$: current}};
    \draw[arr] (s)--(a);

    \node[component] (g) at (2.50,0.45) {GPU};
    \node[component,left=2mm of g] (q) {Scheduler};
    \node[component,left=2mm of q] (o) {Output};
    \node[component,right=2mm of g] (k) {KV};
    \node[component,right=2mm of k] (m) {Memory};
    \coordinate (bus) at (2.50,1.10);
    \draw[merge,draw=orange!65!black] (bus -| o)--(bus -| m);
    \draw[merge,draw=orange!65!black] (a.south)--(bus);
    \draw[fan] (bus -| o)--(o.north);
    \draw[fan] (bus -| q)--(q.north);
    \draw[fan] (bus -| g)--(g.north);
    \draw[fan] (bus -| k)--(k.north);
    \draw[fan] (bus -| m)--(m.north);
  \end{tikzpicture}}
  \caption{The runtime selects the revised context.  One execution-version
  authority then governs its effects across the serving stack.}
  \label{fig:many-to-one}
\end{figure}
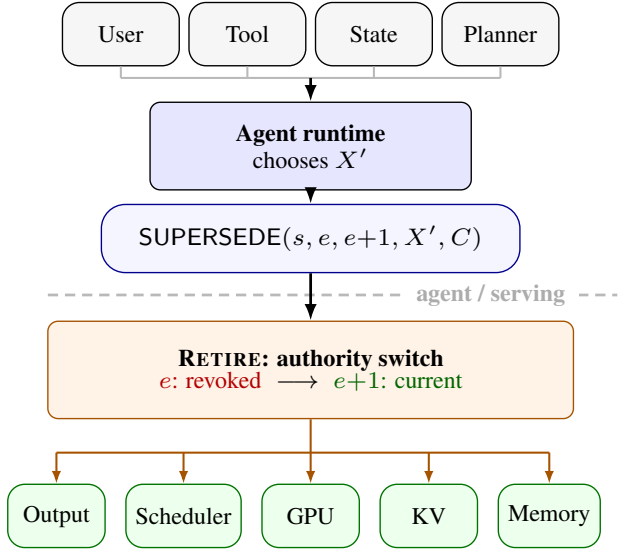

We call the replacement of a live execution
an \emph{agent interruption}.  Figure~\ref{fig:many-to-one} shows the division
of responsibility: the runtime selects what replaces the old execution; the
server enforces that selection across output, computation, and memory.  This
requires a serving abstraction that represents both the authority change and
the state shared across it.

\paragraph{Requests allocate resources; versions determine authority.}
A request tells the server what to schedule and which resources to account for.
An execution version tells it which results may still become visible.  A
\emph{scope} groups the work that the runtime replaces together, including
multiple requests when they belong to the same abandoned plan.  Once that plan
is replaced, its requests can still own buffers and GPU allocations even though
their results have lost authority.  Conversely, their completed common-prefix
state can remain useful to the successor.

The request-centric serving stack couples these roles.  It admits, schedules,
streams, and accounts for KV ownership by request or sequence.  PagedAttention
makes the resource unit concrete: each active sequence has a logical block
table, even when physical blocks are shared \citep{kwon2023pagedattention}.
Aborting a request changes its scheduling status, but the revision also has to
reach output already queued for delivery, GPU work already issued, and state
registered for reuse or transfer.  Caching and tiering extend those objects'
lifetimes and increase the value of coordinating their handoff
\citep{zheng2024sglang,gao2024cachedattention,yu2025pensieve,
gao2025hcache,qin2025mooncake}.

\begin{figure*}[t]
  \centering
  \resizebox{0.98\textwidth}{!}{%
  \begin{tikzpicture}[
      x=1cm,y=1cm,font=\small,
      lane/.style={font=\bfseries,anchor=east},
      current/.style={draw,rounded corners=2pt,fill=blue!10,
                      draw=blue!55!black},
      obsolete/.style={draw,rounded corners=2pt,fill=red!12,
                       draw=red!65!black},
      repeat/.style={draw,rounded corners=2pt,fill=orange!16,
                     draw=orange!70!black},
      inherit/.style={draw,rounded corners=2pt,fill=green!13,
                      draw=green!55!black},
      cleanup/.style={draw,rounded corners=2pt,fill=gray!12,
                      draw=gray!65,dashed},
      axis/.style={-{Latex[length=2mm]},thick,draw=gray!70},
      event/.style={draw=red!70!black,densely dashed,thick},
      note/.style={font=\scriptsize,align=center}
    ]
    \node[lane] at (1.25,1.45) {Native abort};
    \node[lane] at (1.25,-0.10) {\retire};

    \draw[axis] (1.50,1.45)--(15.45,1.45);
    \draw[axis] (1.50,-0.10)--(15.45,-0.10);
    \draw[event] (4.95,-1.22)--(4.95,2.55);
    \node[font=\bfseries,text=red!70!black,anchor=south] at (4.95,2.56)
      {SUPERSEDE $e\!\rightarrow\!e{+}1$};

    \path[current] (1.55,1.12) rectangle (4.85,1.78);
    \node at (3.20,1.45) {authoritative $e$};
    \path[obsolete] (5.05,1.12) rectangle (7.65,1.78);
    \node at (6.35,1.45) {obsolete tail};
    \path[repeat] (7.75,1.12) rectangle (11.25,1.78);
    \node[align=center] at (9.50,1.45) {cold prefill\\valid prefix recomputed};
    \path[current] (11.35,1.12) rectangle (13.65,1.78);
    \node at (12.50,1.45) {$e{+}1$ execution};
    \draw[-{Latex[length=1.6mm]},red!70!black,thick]
      (5.65,1.82)--(5.65,2.14);
    \draw[-{Latex[length=1.6mm]},red!70!black,thick]
      (6.35,1.82)--(6.35,2.14);
    \draw[-{Latex[length=1.6mm]},red!70!black,thick]
      (7.05,1.82)--(7.05,2.14);
    \node[note,text=red!70!black] at (6.35,2.32) {old-version output};
    \node[note,anchor=north] at (6.35,1.04) {predecessor delay};
    \node[note,anchor=north] at (9.50,1.04) {repeated valid work};
    \draw[very thick,blue!70!black] (13.65,1.08)--(13.65,1.82);
    \node[note,anchor=south west] at (13.80,1.58) {first current token};

    \path[current] (1.55,-0.43) rectangle (4.85,0.23);
    \node at (3.20,-0.10) {authoritative $e$};
    \path[obsolete] (5.05,-0.43) rectangle (6.15,0.23);
    \node at (5.60,-0.10) {\textsc{Fence}};
    \path[inherit] (6.25,-0.43) rectangle (8.45,0.23);
    \node[align=center] at (7.35,-0.10) {\textsc{Align}\\certified prefix};
    \path[current] (8.55,-0.43) rectangle (10.85,0.23);
    \node at (9.70,-0.10) {$e{+}1$ execution};
    \path[cleanup] (8.55,-1.10) rectangle (11.25,-0.62);
    \node at (9.90,-0.86) {\textsc{Reclaim} (async)};
    \draw[very thick,blue!70!black] (10.85,-0.47)--(10.85,0.27);
    \node[note,anchor=south west] at (11.00,0.03) {first current token};

    \node[note,anchor=west] at (1.55,-1.50)
      {\tikz\draw[fill=red!12,draw=red!65!black] (0,0) rectangle (0.28,0.18);\ obsolete work};
    \node[note,anchor=west] at (4.00,-1.50)
      {\tikz\draw[fill=orange!16,draw=orange!70!black] (0,0) rectangle (0.28,0.18);\ repeated valid work};
    \node[note,anchor=west] at (7.20,-1.50)
      {\tikz\draw[fill=green!13,draw=green!55!black] (0,0) rectangle (0.28,0.18);\ inherited valid state};
  \end{tikzpicture}}
  \caption{Abort followed by a cold restart pays for both the obsolete tail
  and prefix recomputation.  \retire shortens the tail with FENCE, preserves
  the certified prefix through ALIGN, and reclaims old resources asynchronously.}
  \label{fig:two-sided-tax}
\end{figure*}
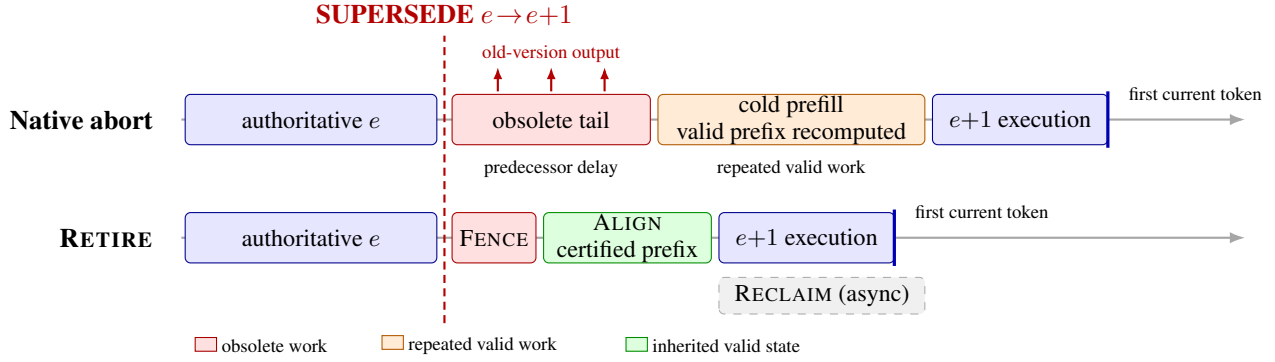

\paragraph{A revision can pay twice.}
Coarse abort-and-restart spends time on work that has become obsolete, then
repeats useful work when the successor starts cold.  Figure~\ref{fig:two-sided-tax}
shows the two delays: waiting for the predecessor to reach a stopping boundary,
and rebuilding the shared prefix.  Faster retirement reduces the first delay;
keeping a valid prefix reduces the second.  Reusing that prefix requires a
writer and ownership boundary so old work cannot subsequently alter it.
The serving system must coordinate cancellation and preservation around the
same version decision.

The consequences appear at multiple boundaries in vLLM.  Across 18 request-abort
trials, consumers receive 170 predecessor tokens in aggregate after
supersession.  A long-prefill replacement waits 3.21 seconds for the next
server boundary.  State handoff adds a third obligation: selecting the completed
prefix whose contents and ownership remain valid.  Together, these observations
motivate a control plane that makes one version decision govern publication,
retirement, and inheritance.

\paragraph{A versioned handoff.}
We present \retire, a redesign of the serving control plane around
\emph{versioned execution}.  \retire first makes obsolete work unable to publish
new effects, then hands its successor only state that satisfies the reuse
contract.  Its organizing rule is \emph{revoke first, preserve second}.
ADVANCE revokes the old version; FENCE bounds its remaining work and closes
writers; ALIGN certifies the reusable prefix; RESUME launches the successor;
and RECLAIM releases isolated old resources asynchronously.  Reclamation stays
off the successor's critical path because those resources no longer determine
its output or state.

We implement this authority boundary throughout vLLM, from output publication
and GPU execution to KV installation, state transfer, and reclamation.
The evaluation first establishes correct publication, inheritance, and tenant
isolation.  Combining fast invalidation with certified inheritance then lowers
paired revision-to-successor TTFT by a median 17.1\% over native abort and cold
restart.  Under intervening HBM churn, certified retention accelerates
post-churn recovery by up to 6.87$\times$/5.07$\times$ on H100 PCIe/GB200.

Finally, replaying recorded coding-agent interruption bursts, \retire publishes
no obsolete output, advances every final version, and delivers 5.53$\times$
the current-version output per second of the best tested native quantum.
Distributed and co-tenant experiments extend the same boundary across workers
and shared resources.  Versioned execution supports both faster handoff and
continued progress when another revision arrives before cleanup ends.

This paper makes three contributions:

\begin{itemize}[leftmargin=*,itemsep=2pt,topsep=3pt]
  \item \textbf{Workload abstraction.} We define revisable execution through
  version transitions and characterize their pressure, retirement horizon,
  reusable state, and scope width.
  \item \textbf{Serving redesign.} We separate request resource ownership from
  execution-version authority and coordinate fast invalidation, certified
  inheritance, and asynchronous reclamation in \retire.
  \item \textbf{Implementation and evidence.} We realize the design across the
  vLLM serving lifecycle and establish correctness, faster recovery, and
  sustained progress under recorded bursts, distributed execution, and tenant
  sharing.
\end{itemize}

\section{Converging Trends and Related Work}
\label{sec:related}

Agent control, stateful serving, and KV management are converging on executions
that change while their work and state remain live.  Together, they motivate
a serving abstraction that coordinates which work loses authority and which
state remains useful.

\paragraph{Agents revise live work.}
Benchmarks and control studies treat changing instructions, goals, and
stop decisions as recurring execution events
\citep{wu2025reasoninginterruptible,rana2025agentchangebench,
zou2026interruptbench,zhai2026revisable,khan2026stop}.  They establish revision
as a workload property: after new information arrives, the runtime may select
a replacement for an already-running execution.  The resulting decision must
reach the serving stack while old computation and output are still in flight.

\paragraph{Servers carry work across time.}
FastServe provides iteration-level preemptive scheduling
\citep{wu2026fastserve}; InferCept preserves state across intercepts
\citep{abhyankar2024infercept}; Agentix schedules agent programs
\citep{luo2026agentix}.  Their scheduling decisions span turns, pauses, and
persistent state.  Preemption retains the identity of paused work; revision is
the complementary transition in which that identity loses authority and a
different successor may take over.

\paragraph{State outlives individual requests.}
PagedAttention and SGLang provide block and prefix sharing
\citep{kwon2023pagedattention,zheng2024sglang}.  CachedAttention, Pensieve,
HCache, and Mooncake retain or restore conversational state across memory tiers
\citep{gao2024cachedattention,yu2025pensieve,gao2025hcache,qin2025mooncake},
while Leyline proposes agentic KV directives \citep{ma2026leyline}.  This
decoupling makes state preservation valuable during revision, but matching
tokens or bytes alone cannot establish whether an in-flight writer or transfer
still belongs to the selected execution.  Rollback consistency exposes the
same need to bound admissible retained state \citep{zhang2026aborted}.

Versioned execution connects these developments through one authority boundary.
The runtime's replacement decision governs publication and retirement, and
determines which retained state the successor may adopt.  \retire unifies
revocation, writer closure, state handoff, and asynchronous reclamation in one
lifecycle.

\section{Revisable Execution as a Serving Workload}
\label{sec:exactness}

The coding-agent example has two execution versions: the abandoned API change
and the revised implementation change.  The runtime chooses between them;
the server must make that choice effective for every affected request and
state object.  We define the serving event around this transfer of authority.

\begin{quote}
\textbf{Definition.} An agent interruption at the serving boundary replaces
a runtime-selected scope at execution version $e$ with version $e{+}1$.
Version $e$ loses permission to publish output or install state; version
$e{+}1$ runs from the revised context supplied by the runtime.
\end{quote}

\subsection{The runtime selects the replacement}

A \emph{scope} is the work replaced together.  It can contain one model
request, several requests for an abandoned branch, or their distributed
workers.  An \emph{execution version} is one attempt to execute that scope.
Other scopes in the same batch keep their own authority.  The runtime submits
\[
  \mathsf{SUPERSEDE}(s,e,e{+}1,X',C),
\]
where $s$ identifies the scope, $e$ is its current version, $e{+}1$ is the
replacement, and $X'$ is the revised context.  The \emph{execution contract}
$C$ specifies when inherited state is acceptable, including the numerical mode
and required successor-output equivalence against clean execution.  After the
switch, $e{+}1$ owns publication authority and can adopt certified predecessor
state.  A pause retains the same authority; a terminal cancellation has no
successor.

\subsection{A taxonomy of agent interruptions}

Table~\ref{tab:taxonomy} gives representative reasons for replacing live work.
A failed tool call, for example, induces a transition when the runtime abandons
work dependent on that call and selects a new plan.  The server receives the
replacement decision and revised context, regardless of the original trigger.
Appendix~\ref{app:taxonomy} expands this mapping.

\begin{table}[t]
  \caption{Representative interruptions.  Every runtime decision shown here
  becomes the same serving-level version transition.}
  \label{tab:taxonomy}
  \centering
  \footnotesize
  \setlength{\tabcolsep}{3pt}
  \begin{tabularx}{\columnwidth}{@{}>{\raggedright\arraybackslash}p{0.49\columnwidth}>{\raggedright\arraybackslash}X@{}}
    \toprule
    \textbf{Source and example trigger} & \textbf{Runtime decision} \\
    \midrule
    \textbf{User steering}: correction or new request &
    Revise goal, context, or branch \\
    \hdashline
    \textbf{Tool/service}: failed call or expired credential &
    Replace dependent work and retry \\
    \hdashline
    \textbf{Environment/validation}: stale state or rejected result &
    Re-observe, repair, or roll back \\
    \hdashline
    \textbf{Planning/operation}: winning branch or changed deadline &
    Switch branch or execution policy \\
    \bottomrule
  \end{tabularx}
\end{table}

\subsection{The workload exposed to the server}

The same transition can impose very different costs.  Four dimensions expose
those costs to the server:

\begin{description}[leftmargin=0pt,labelindent=0pt,itemsep=2pt]
  \item[Revision pressure:] arrival rate and burstiness.  Another correction
  may arrive before the previous version finishes draining.
  \item[Retirement horizon:] outstanding work before a safe boundary.  A long
  prefill or speculative step can delay its replacement.
  \item[Successor overlap:] state shared with the revised context.  A retained
  repository prefix can save most of a successor's prefill.
  \item[Scope width:] requests, cache groups, transfers, and workers replaced
  together.  A distributed execution requires coordinated closure.
\end{description}

These dimensions determine the value of finer execution boundaries, retained
state, and coordinated retirement.  Across them, a successful transition has
three outcomes:

\begin{enumerate}[leftmargin=*,itemsep=1pt,topsep=2pt]
  \item \textbf{Current output:} revoked work cannot publish new effects.
  \item \textbf{Valid inheritance:} the successor receives only state that
  satisfies its execution contract.
  \item \textbf{Efficient handoff:} useful work survives the switch, obsolete
  work leaves its critical path, and unrelated tenants remain isolated.
\end{enumerate}

The next section follows these obligations through a request-centric server
to show why resource lifetime and execution authority need separate units.

\section{From Request-Centric to Versioned Serving}
\label{sec:design}

Return to the coding agent when the user changes the plan.  The runtime sends
an abort for the model request representing the old plan.  The scheduler can
remove its future work, but a token batch may already be waiting for a slow
consumer, a GPU step may still be processing the old context, and completed
repository-prefix blocks may be registered in the cache.  All belong to the
abandoned execution, yet they require different actions: suppress the output,
bound the computation, and preserve the useful prefix.

A request ID identifies resources; a version determines whether their output
or state may still affect the user or the successor.
Table~\ref{tab:cancel-vs-transition} follows the decision through the stack.
Versioned serving makes the distinction explicit: requests own schedulable
work and allocations, while the current scope version determines which
effects can become visible.  A replacement can then revoke old publication
before all old resources are released, and inherit useful state without
keeping the predecessor authoritative.

\begin{table*}[t]
  \caption{A request is a useful resource unit but an insufficient authority
  unit once an execution scope has been replaced.}
  \label{tab:cancel-vs-transition}
  \centering
  \small
  \setlength{\tabcolsep}{5pt}
  \begin{tabularx}{\textwidth}{@{}p{0.17\textwidth}p{0.34\textwidth}X@{}}
    \toprule
    \textbf{Moment} & \textbf{Request/resource view} &
    \textbf{Scope/version authority view} \\
    \midrule
    Runtime decision &
    Cancel one or more request IDs &
    Revoke the old scope version before any later publication \\
    Buffered output &
    Previously produced frames remain ordinary results &
    Frames from the revoked version are no longer publishable \\
    Issued GPU work &
    Runs until the next request boundary &
    Is bounded, accounted for, and unable to commit stale state \\
    Cached or tiered KV &
    Matching tokens or bytes identify a candidate &
    Reuse also closes old writers and validates configuration and generation \\
    Physical cleanup &
    Request lifetime determines release &
    Cleanup follows the last old writer and need not delay the successor \\
    \bottomrule
  \end{tabularx}
\end{table*}

\paragraph{Two costs on one critical path.}
For a coarse abort followed by cold prefill, the dominant sequential costs are
the predecessor's remaining time to a server-visible boundary,
$T_{\mathit{tail}}$, and the successor's full prefill,
$T_{\mathit{full}}$:
\[
  T_{\mathit{abort}} = T_{\mathit{tail}} + T_{\mathit{full}}.
\]
A versioned transition can replace that tail with time to a bounded safe point,
$T_{\mathit{safe}}$, certify inherited state in $T_{\mathit{align}}$, and
prefill the remaining suffix in $T_{\mathit{suffix}}$:
\[
  T_{\mathit{versioned}} = T_{\mathit{safe}} + T_{\mathit{align}}
                          + T_{\mathit{suffix}}.
\]
These decompositions expose the two levers: shorten obsolete work and avoid
repeating valid work.  When a native prefix hit already removes recomputation,
the remaining opportunity lies in retirement and coordination.  To combine
the levers, the server must control the following three boundaries.

\subsection{Obsolete results can escape}

Removing a request from the scheduler does not retract tokens already queued
between a worker, EngineCore, and a slow consumer.  Request completion and API
publication are different moments.  The server must check the current execution
version again when delivering output, so a buffered result from an abandoned
plan cannot reach the application after the switch.

\subsection{Obsolete work can delay the replacement}

Host-side cancellation cannot retract work already submitted to a model
worker.  A long prefill, speculative step, or distributed group may continue
until the next boundary recognized by the server.  Physical completion and
logical invalidation have different deadlines: old work may finish after a
revision, but its output and state commits must lose authority immediately.
The successor needs writer-closed shared state and isolation from old-only
state.  It need not wait for every predecessor allocation to be reclaimed.

\subsection{State handoff lacks versioned ownership}

The common prompt prefix identifies candidate state for reuse.  A prefix
cache can make its completed blocks available, while the transition needs
a stable handoff: close predecessor writers, retain the blocks until adoption,
and validate their ownership generation.  A tiered transfer must likewise be
installed for the intended successor.  Token positions, model and adapter
configuration, cache layout, tenant, and execution mode determine which state
is compatible.  Rollback consistency makes the importance of bounding retained
state explicit \citep{zhang2026aborted}.

The server therefore needs both an eligibility decision and a lifetime
decision: which state may be inherited, and how it remains available until
the successor takes ownership.  A completed common prefix can be retained;
an unfinished or incompatible suffix must be recomputed.  Releasing the old
request and later looking for a cache hit leaves this handoff opportunistic:
intervening requests may evict the blocks before the successor arrives.

\retire connects these boundaries through one version lifecycle.  Revocation
decides what may publish; writer closure makes inheritance safe; certified
handoff preserves useful work; and isolation lets reclamation proceed after
the successor starts.  Together they turn the two performance levers into a
coordinated serving transition.

\section{\retire: A Versioned Serving Control Plane}
\label{sec:scheduler}

\retire organizes serving around a simple rule: \emph{revoke first, preserve
second}.  The old plan immediately loses authority to publish, while its useful
state survives through an explicit handoff.  Resources are released after
their last users finish; only state the successor will consume must be ready
before it starts.

\retire assigns authority to $(scope,version)$, with one current version per
runtime-selected scope.  The implementation represents the version as a
monotonically increasing \emph{epoch}.  Each request, output frame, model step,
and transfer carries that identity, and publication checks it against the
scope's current epoch.  KV records also carry a \emph{generation} identifying
an allocation's ownership instance.  Installation checks that generation;
writer fences and delayed recycling protect the underlying pages.

\subsection{Three guarantees across the serving lifecycle}

\retire enforces three rules at the serving boundaries where state becomes
visible:

\begin{description}[leftmargin=0pt,labelindent=0pt,itemsep=2pt]
  \item[Current-version publication.] Tokens, scheduler results, and installed
  state are visible only if they belong to the current scope epoch.
  \item[Versioned handoff.] Predecessor state is consumed only when its token
  prefix and execution configuration match, its old writer is closed, and its
  ownership generation remains valid.  Otherwise the successor recomputes.
  \item[Retired-state isolation.] Old work may finish, but it cannot overwrite
  state owned by the successor or an unrelated tenant.
\end{description}

Publication and isolation establish the switch; versioned handoff recovers
performance through safe inheritance.  The execution contract selects the
required output comparison, including batch-invariant execution when numerical
equivalence between cold and reused computation is required.

The two performance mechanisms share the same authority boundary.  FENCE
shortens the predecessor tail and closes writers; ALIGN uses that closure to
separate the committed common prefix from the obsolete suffix.  If no reuse is
profitable, a cold successor bypasses physical pinning and starts under an
independent cache generation.  If reuse is profitable, only certified pages
cross the version boundary.  The transition chooses its state path after
establishing which work remains useful.

\subsection{One five-stage transition}

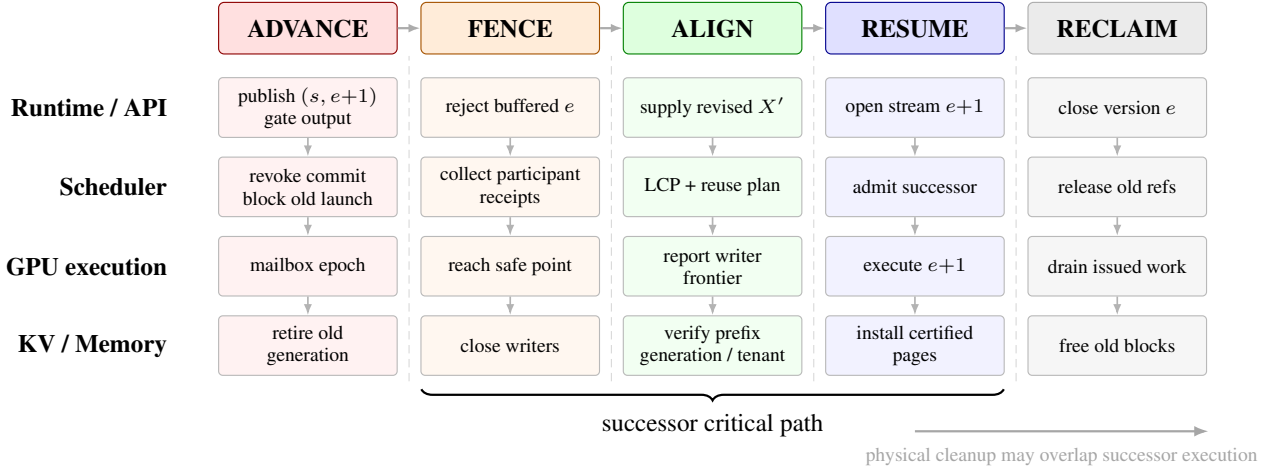
\begin{figure*}[t]
  \centering
  \resizebox{0.98\textwidth}{!}{%
  \begin{tikzpicture}[
      x=1cm,y=1cm,font=\scriptsize,
      lane/.style={font=\small\bfseries,anchor=east,align=right},
      header/.style={draw,rounded corners=2pt,minimum width=2.25cm,
                     minimum height=6.5mm,font=\small\bfseries},
      cell/.style={draw=gray!55,rounded corners=1.5pt,minimum width=2.25cm,
                   minimum height=7.5mm,align=center,inner xsep=2pt},
      flow/.style={-{Latex[length=1.5mm]},draw=gray!65,thick},
      down/.style={-{Latex[length=1.4mm]},draw=gray!60,semithick}
    ]
    \node[lane] at (1.90,2.05) {Runtime / API};
    \node[lane] at (1.90,1.05) {Scheduler};
    \node[lane] at (1.90,0.05) {GPU execution};
    \node[lane] at (1.90,-0.95) {KV / Memory};

    \node[header,fill=red!12,draw=red!55!black] (ha) at (3.55,3.05) {ADVANCE};
    \node[header,fill=orange!15,draw=orange!65!black] (hf) at (6.10,3.05) {FENCE};
    \node[header,fill=green!13,draw=green!50!black] (hl) at (8.65,3.05) {ALIGN};
    \node[header,fill=blue!12,draw=blue!55!black] (hr) at (11.20,3.05) {RESUME};
    \node[header,fill=gray!16,draw=gray!65] (hc) at (13.75,3.05) {RECLAIM};
    \draw[flow] (ha.east)--(hf.west);
    \draw[flow] (hf.east)--(hl.west);
    \draw[flow] (hl.east)--(hr.west);
    \draw[flow] (hr.east)--(hc.west);

    \foreach \x in {4.825,7.375,9.925,12.475}
      \draw[densely dashed,gray!30] (\x,-1.40)--(\x,2.70);

    \node[cell,fill=red!5] (a1) at (3.55,2.05)
      {publish $(s,e{+}1)$\\gate output};
    \node[cell,fill=red!5] (a2) at (3.55,1.05)
      {revoke commit\\block old launch};
    \node[cell,fill=red!5] (a3) at (3.55,0.05)
      {mailbox epoch};
    \node[cell,fill=red!5] (a4) at (3.55,-0.95)
      {retire old\\generation};

    \node[cell,fill=orange!6] (f1) at (6.10,2.05)
      {reject buffered $e$};
    \node[cell,fill=orange!6] (f2) at (6.10,1.05)
      {collect participant\\receipts};
    \node[cell,fill=orange!6] (f3) at (6.10,0.05)
      {reach safe point};
    \node[cell,fill=orange!6] (f4) at (6.10,-0.95)
      {close writers};

    \node[cell,fill=green!5] (l1) at (8.65,2.05)
      {supply revised $X'$};
    \node[cell,fill=green!5] (l2) at (8.65,1.05)
      {LCP + reuse plan};
    \node[cell,fill=green!5] (l3) at (8.65,0.05)
      {report writer\\frontier};
    \node[cell,fill=green!5] (l4) at (8.65,-0.95)
      {verify prefix\\generation / tenant};

    \node[cell,fill=blue!5] (r1) at (11.20,2.05)
      {open stream $e{+}1$};
    \node[cell,fill=blue!5] (r2) at (11.20,1.05)
      {admit successor};
    \node[cell,fill=blue!5] (r3) at (11.20,0.05)
      {execute $e{+}1$};
    \node[cell,fill=blue!5] (r4) at (11.20,-0.95)
      {install certified\\pages};

    \node[cell,fill=gray!7] (c1) at (13.75,2.05)
      {close version $e$};
    \node[cell,fill=gray!7] (c2) at (13.75,1.05)
      {release old refs};
    \node[cell,fill=gray!7] (c3) at (13.75,0.05)
      {drain issued work};
    \node[cell,fill=gray!7] (c4) at (13.75,-0.95)
      {free old blocks};

    \foreach \top/\mid/\low/\bottom in
      {a1/a2/a3/a4,f1/f2/f3/f4,l1/l2/l3/l4,r1/r2/r3/r4,c1/c2/c3/c4} {
      \draw[down] (\top.south)--(\mid.north);
      \draw[down] (\mid.south)--(\low.north);
      \draw[down] (\low.south)--(\bottom.north);
    }

    \draw[decorate,decoration={brace,mirror,amplitude=4pt},thick]
      (4.98,-1.50)--(12.32,-1.50)
      node[midway,below=5pt,font=\small]{successor critical path};
    \draw[-{Latex[length=1.8mm]},gray!70,thick]
      (11.20,-2.03)--(14.90,-2.03)
      node[midway,below=2pt,font=\scriptsize]{physical cleanup may overlap successor execution};
  \end{tikzpicture}}
  \caption{\retire is one cross-layer version transition: ADVANCE changes
  authority, FENCE closes old publication and writers, ALIGN certifies inherited
  state, RESUME opens the successor, and RECLAIM releases isolated state off the
  critical path.}
  \label{fig:path}
\end{figure*}

Figure~\ref{fig:path} shows how the decision reaches the serving stack:

\begin{description}[leftmargin=0pt,labelindent=0pt,itemsep=2pt]
  \item[ADVANCE changes authority.] Publish version $e{+}1$ for the scope.
  Queued output and new commits carrying $e$ are rejected.
  \item[FENCE bounds obsolete work.] Stop new old-version launches and close
  writers at supported safe boundaries, collecting participant receipts.
  \item[ALIGN selects useful state.] Match the revised context to the
  completed prefix and certify its configuration, generation, and ownership.
  \item[RESUME starts the replacement.] Install the certified prefix, allocate
  the new suffix, and admit work under $e{+}1$.
  \item[RECLAIM releases isolated resources.] After their final users drain,
  free old references and allocations asynchronously.
\end{description}

The stages express dependencies: revoke before publication, close writers
before inheritance, and isolate state before recycling it.  Successor progress
depends on the state it will consume, rather than on the release of every
predecessor allocation.

\subsection{Resume before reclamation}

In the coding-agent example, the successor adopts the certified repository
prefix and allocates fresh pages for its changed suffix.  Old suffix pages and
buffered results are unreachable from that successor.  Their owners can drain
while the revised plan runs: old publication is revoked, and old pages remain
isolated until their writers finish.

The common prefix stays pinned until the successor adopts it.  Once adoption
is verified, handoff pins are released and the successor's references keep
those pages live.  Only immutable, writer-closed pages cross this boundary,
allowing preservation and cleanup to proceed under the same authority decision.

\section{A Versioned vLLM Serving Stack}
\label{sec:implementation}

We implement \retire in vLLM 0.13.0 by carrying $(scope,epoch)$ through request
admission, scheduling, model execution, output publication, KV installation,
and memory transfer.  Request IDs index local resources; the scope epoch
determines whether their results can become visible.

Three changes enforce the lifecycle: publication gates reject obsolete output,
model safe points bound obsolete computation, and versioned KV ownership
controls which state the successor adopts.

\subsection{Redesigning the execution lifecycle}

\paragraph{Version-governed publication.}
Every native request maps to $(scope,epoch)$.  The server checks that authority
at scheduling and model completion, when output enters the streaming queue,
and again when the API consumer receives it.  The delivery check rejects a
frame whose version was superseded while it waited in the queue.

\paragraph{Bounded retirement inside model execution.}
An epoch update reaches workers outside ordinary request traffic.  A
\emph{safe point} is a model boundary where the worker can stop launching
additional old-version work.  Writer receipts establish the completed-state
frontier, while KV and output gates reject stale commits.  Short decode steps
already reach model-step boundaries frequently.  Long steps arm additional
device predicates and bounded checks, avoiding per-layer checking costs on
short steps while shortening the tail of a long prefill.

\paragraph{Versioned KV-cache transition.}
FENCE snapshots the predecessor block table before releasing its request and
pins its completed prefix.  ALIGN finds the longest common token prefix,
limits it to the minimum writer-complete frontier across participants, and
rounds it to a block boundary shared by every KV group.  It then checks model
and cache configuration, tenant, generation, transfer identity, and numerical
mode.  This produces the only prefix that may cross the version boundary.

\begin{figure}[t]
  \centering
  \resizebox{0.98\columnwidth}{!}{%
  \begin{tikzpicture}[
      font=\scriptsize,
      label/.style={font=\small\bfseries,anchor=east},
      block/.style={draw,rounded corners=1.5pt,minimum width=7.5mm,
                    minimum height=6.5mm,inner sep=0pt},
      shared/.style={block,fill=green!13,draw=green!50!black},
      oldonly/.style={block,fill=red!11,draw=red!60!black},
      newonly/.style={block,fill=blue!11,draw=blue!55!black},
      keep/.style={{Latex[length=1.1mm]}-{Latex[length=1.1mm]},
                   draw=green!50!black,thick},
      drop/.style={-{Latex[length=1.4mm]},draw=red!60!black,dashed,thick}
    ]
    \node[label] at (1.00,1.28) {old $e$};
    \node[label] at (1.00,0.18) {new $e{+}1$};
    \foreach \x/\n in {1.50/A,2.35/B,3.20/C} {
      \node[shared] (o\n) at (\x,1.28) {\n};
      \node[shared] (n\n) at (\x,0.18) {\n};
    }
    \node[oldonly] (oD) at (4.05,1.28) {D};
    \node[oldonly] (oE) at (4.90,1.28) {E};
    \node[newonly] (nF) at (4.05,0.18) {F};
    \node[newonly] (nG) at (4.90,0.18) {G};
    \node[font=\scriptsize\bfseries,text=green!40!black,anchor=south]
      at (2.35,1.72) {certified prefix};
    \node[font=\scriptsize\bfseries,text=red!65!black,anchor=south]
      at (4.48,1.72) {old-only suffix};
    \node[font=\scriptsize\bfseries,text=blue!60!black,anchor=north]
      at (4.48,-0.18) {new-only suffix};
    \node[draw=gray!60,rounded corners=2pt,fill=gray!8,
          minimum height=6.5mm] (gc) at (6.30,1.28) {async reclaim};
    \draw[drop] (oE.east)--(gc.west);
    \draw[keep] (oB.south)--node[right=0.7mm,fill=white,inner sep=0.4pt,
      font=\scriptsize,text=green!35!black] {alias} (nB.north);
  \end{tikzpicture}}
  \caption{The successor table aliases the certified physical HBM prefix,
  allocates a new suffix, and leaves the unreachable old suffix to reclamation.}
  \label{fig:kv-transition}
\end{figure}
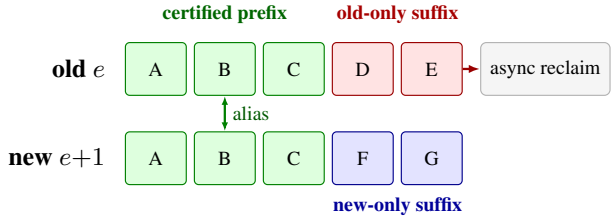

At RESUME, the successor table aliases those physical HBM pages and allocates
its changed suffix under $e{+}1$.  The old suffix loses cache reachability and
is reclaimed after its last writer; the shared prefix stays pinned through
adoption (Figure~\ref{fig:kv-transition}).  Keeping this state in place avoids
copying, eviction, restoration, and recomputation.  When
HBM capacity must be returned before the successor arrives, the same
certificate governs DRAM offload and restoration.  A failed certificate selects
a cold, fresh-generation successor.  A restored object is installed only if
its target version and generation still match, rejecting transfers that
complete after their intended execution has been retired.

The contract extends across speculation, multiple KV groups, memory tiers,
tenants, and distributed participants; their adapters and failure evidence
appear in Appendix~\ref{app:details}.

\setlength{\textfloatsep}{10pt}
\setlength{\dbltextfloatsep}{10pt}
\section{Evaluation}
\label{sec:evaluation}

The evaluation follows the handoff from its guarantees to its benefits.
First, does the successor receive valid state while obsolete effects are
suppressed?  Second, how much time is saved by shortening old work and
preserving useful state, separately and together?  Finally, can the same
transition sustain progress through recorded interruption bursts, distributed
execution, and tenant sharing?  Performance results come from runs that pass
the corresponding correctness checks.

\subsection{Experimental method}

Integrated runs use vLLM with H100 PCIe GPUs for the main and distributed
paths; the state-handoff campaign repeats the same design on H100 PCIe and
GB200.  Qwen3-32B in BF16 is the primary configuration; GPT-OSS-20B supplies
heterogeneous KV groups and trace replay.  Single-GPU retirement paths use
three pairs, active speculative decode and distributed topologies use five,
and state handoff uses four paired quartets at each of nine pressure points per
platform.  Co-tenant isolation uses three seeds.

Each pair shares a predecessor, revision, and execution mode.  The checker
compares current-version output, KV ownership and transfer digests, and
distributed receipts against a clean reference before admitting performance
data.  We report medians and compute speedups before display rounding.
Appendix~\ref{app:details} gives the complete configurations and trace construction.

\paragraph{Comparisons and metrics.}
Table~\ref{tab:eval-paths} identifies the mechanisms enabled in each comparison.
\emph{Single-handoff TTFT} spans the version-switch signal to the first
authoritative successor token.  \emph{FENCE latency} isolates the predecessor's
writer closure.  \emph{State-recovery TTFT} starts after intervening HBM churn
and ends at that successor token.  \emph{Current-version goodput} counts
authoritative tokens per second over the replay window.  Thus the composition
measures a complete handoff, the retirement and recovery measurements explain
its costs, and the burst replay measures sustained useful output.

\begin{table}[t]
  \caption{Baseline and versioned state paths.  The composition experiment
  separately enables invalidation and inheritance.}
  \label{tab:eval-paths}
  \centering
  \footnotesize
  \setlength{\tabcolsep}{3pt}
  \begin{tabularx}{\columnwidth}{@{}>{\raggedright\arraybackslash}p{0.26\columnwidth}>{\raggedright\arraybackslash}X@{}}
    \toprule
    \textbf{Path} & \textbf{Enabled mechanisms} \\
    \midrule
    Native+cold & Request abort; cold successor prefill \\
    Native+prefix & Tuned abort and chunked prefill; reuse available native-cache blocks \\
    \retire HBM & Versioned handoff; pin certified blocks until successor adoption \\
    \retire DRAM & Versioned handoff; offload certified state, release HBM, then restore \\
    \bottomrule
  \end{tabularx}
\end{table}

\subsection{Correctness: closing the version lifecycle}
\label{sec:eval-correctness}

We first test the two obligations created by the coding-agent example:
discard the old plan's effects and preserve only valid context.  Across 18
native publication trials, consumers receive 170 predecessor tokens after
abort because those frames have crossed internal queues.  \retire's
publication boundary suppresses them.  For inherited state, we hold the
successor, reuse length, and execution shape fixed while changing the source
of its pages.  Completed-predecessor and clean-prefix pages agree in all ten
seed--LCP configurations, controlling for batch-dependent numerical variation.  An injected
page-content fault leaves the cache count and physical IDs unchanged but
changes the successor in 3/3 trials; restoring the certified bytes recovers
exact output in 3/3.

The publication control tests revocation across internal queues; the page-fault
control tests inherited contents with block identity and execution shape held
fixed.  Together they establish the output and state boundaries of a valid
handoff.

We then cross buffered output, long prefill, draft--verify prefill, active
speculative decode, two heterogeneous KV groups, and co-tenant execution with
one version transition.  These cases cover publication, scheduler commit, KV
installation, and independently retiring participants.

\begin{strip}
  \centering
  \captionof{table}{Correctness closure across the version lifecycle.  Successor
  comparisons use the declared clean execution mode; the faulted-page row is
  an injected positive control.}
  \label{tab:correctness}
  \small
  \setlength{\tabcolsep}{4pt}
  \begin{tabularx}{\textwidth}{@{}>{\raggedright\arraybackslash}p{0.17\textwidth}>{\raggedright\arraybackslash}p{0.29\textwidth}>{\raggedright\arraybackslash}p{0.27\textwidth}>{\raggedright\arraybackslash}X@{}}
    \toprule
    \textbf{Lifecycle surface} & \textbf{Request-local behavior} &
    \textbf{\retire closure} & \textbf{Observed result} \\
    \midrule
    Output publication &
    170 obsolete tokens across 18 trials &
    Gate API dequeue by current authority &
    Zero predecessor output \\
    Long prefill &
    Old work drains to request boundary &
    Reject stale launch, KV, and scheduler commit &
    3/3 successors exact \\
    Draft--verify / active decode &
    Old speculative work remains live &
    Fence draft and target work under one version &
    3/3 and 5/5 exact \\
    Two KV groups &
    Independent participant state &
    Fence both groups at one authority coordinate &
    3/3 successors exact \\
    Faulted aligned pages &
    Same cache count and block IDs; 3/3 diverge &
    Restore certified bytes or reject reuse &
    3/3 successors exact \\
    Tenant isolation &
    A changes while B decodes &
    Scope authority and cache ownership by tenant &
    A exact; B's audited KV unchanged in 3/3 \\
    \bottomrule
  \end{tabularx}
\end{strip}

Table~\ref{tab:correctness} follows the authority boundary through these paths.
In every integrated prefill,
speculative, and heterogeneous-cache trial, no predecessor output is published
and the successor matches its clean reference.  The active speculative case
waits until draft work is observed before issuing SUPERSEDE, so the transition
occurs inside speculation.  The publication and commit gates also reject late
completions independently of when the native request object disappears.

Output, page digests, ownership generations, and participant receipts thus
validate one authority boundary across the serving lifecycle.  These accepted
runs provide the basis for measuring its performance.

\subsection{Performance: removing both transition costs}
\label{sec:eval-performance}

Once correctness is established, the performance question has two parts.
Native replacement can wait for the obsolete predecessor to reach a request
boundary and then rebuild state that remains valid for the successor.
FENCE shortens the obsolete tail; ALIGN and RESUME preserve the valid prefix.

\paragraph{Combined handoff latency.}
We first measure both mechanisms on the same revision-to-successor critical
path: native abort plus cold restart, fast invalidation alone, certified
inheritance alone, and both mechanisms together.

\begingroup
\setlength{\intextsep}{4pt}
\begin{table}[H]
  \centering
  \small
  \caption{Fast invalidation and inheritance compose.  TTFT is median
  [min, max]; reductions are medians of paired changes against native abort
  plus cold restart.}
  \label{tab:composition}
  \setlength{\tabcolsep}{3.3pt}
  \begin{tabular}{lrrr}
    \toprule
    Arm & TTFT (ms) & Paired reduction & Reuse \\
    \midrule
    Native+cold & 1127 [1124,1178] & --- & 0 \\
    Invalidate & 1068 [1064,1114] & 5.6\% & 0 \\
    Inherit & 1011 [1004,1033] & 10.3\% & 4096 \\
    Combined & \textbf{951 [928,976]} & \textbf{17.1\%} & 4096 \\
    \bottomrule
  \end{tabular}
\end{table}
\endgroup

Table~\ref{tab:composition} uses an 8192-token prompt, a 4096-token prefill
quantum, 90\% overlap, and a revision 1.1 seconds into prefill.  The combined
path reduces paired successor TTFT by 17.1\% (15.4--17.6\%) and adopts 4096
certified tokens in every trial.  The two mechanisms address different terms:
invalidation shortens the predecessor tail; inheritance shortens successor
reconstruction.  Their combination exceeds either mechanism alone.

\paragraph{Where the time is saved.}
The obsolete tail grows with execution granularity: 4K, 8K, and 16K active
prefill units take 714.9, 1541.2, and 3886.7~ms to return to the ordinary
EngineCore boundary.  Prefix reconstruction grows with shared context: a
successor with a certified 4080-token prefix takes approximately 3965~ms to prefill cold in
batch-invariant mode and 117~ms when it inherits that prefix.  This isolated
inheritance experiment reduces successor TTFT by 33.8$\times$.

Fast retirement generalizes across execution structures.  FENCE improves by
24.9$\times$ for long prefill, 51.3$\times$ for draft--verify prefill,
9.99$\times$ during active speculative decode, and 7.01$\times$ across two
heterogeneous KV groups.  The corresponding interruption-to-successor gains
are 1.12--1.48$\times$ after including useful successor computation.  For long
prefill, FENCE falls from 3213 to 129~ms and total transition time from 14.6 to
9.86 seconds.  Appendix~\ref{tab:supp-retire} gives every path.

\paragraph{State availability under pressure.}
A full-hit control gives native prefix caching every reusable block, submits the
successor immediately, and introduces no HBM pressure.  Across both GPUs and
three interruption offsets, versioned handoff stays within 2\% of that strong
baseline.  We next increase the state churn between interruption and successor
admission.

\begingroup
\setlength{\intextsep}{4pt}
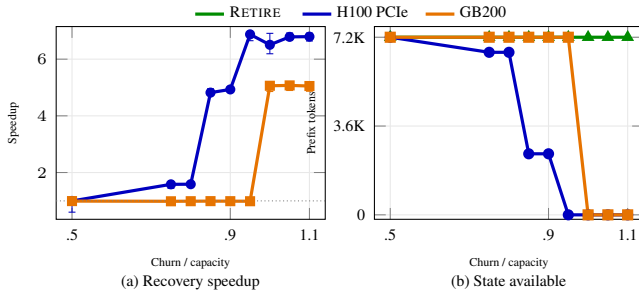
\begin{figure}[H]
  \centering
  {\fontsize{6}{7}\selectfont
    \tikz[baseline=-0.45ex]{\draw[very thick,green!55!black] (0,0)--(0.38,0);}
    \retire\quad
    \tikz[baseline=-0.45ex]{\draw[very thick,blue!75!black] (0,0)--(0.38,0);}
    H100 PCIe\quad
    \tikz[baseline=-0.45ex]{\draw[very thick,orange!90!black] (0,0)--(0.38,0);}
    GB200\par}
  \vspace{1pt}
  \begin{minipage}[t]{0.49\linewidth}
    \centering
    \begin{tikzpicture}[trim axis left,trim axis right]
      \begin{axis}[
        scale only axis,width=0.88\linewidth,height=2.55cm,
        xmin=0.46,xmax=1.14,ymin=0.35,ymax=7.15,
        xtick={0.5,0.9,1.1},
        xticklabels={.5,.9,1.1},
        xlabel={Churn / capacity},ylabel={Speedup},
        tick label style={font=\tiny},
        label style={font=\fontsize{4.5}{5}\selectfont},
        grid=major,grid style={gray!18},
      ]
        \addplot[very thick,blue!75!black,mark=*,mark size=1.3pt,
          error bars/.cd,y dir=both,y explicit,error bar style={line width=0.35pt}]
          table[x=churn,y=retention,y error minus=ret_err_minus,
                y error plus=ret_err_plus] {figures/state_handoff_h100.dat};
        \addplot[very thick,orange!90!black,mark=square*,mark size=1.3pt,
          error bars/.cd,y dir=both,y explicit,error bar style={line width=0.35pt}]
          table[x=churn,y=retention,y error minus=ret_err_minus,
                y error plus=ret_err_plus] {figures/state_handoff_gb200.dat};
        \addplot[thin,densely dotted,gray] coordinates {(0.46,1) (1.14,1)};
      \end{axis}
    \end{tikzpicture}
    \par\vspace{-4pt}\tiny (a) Recovery speedup
  \end{minipage}\hfill
  \begin{minipage}[t]{0.49\linewidth}
    \centering
    \begin{tikzpicture}[trim axis left,trim axis right]
      \begin{axis}[
        scale only axis,width=0.88\linewidth,height=2.55cm,
        xmin=0.46,xmax=1.14,ymin=-180,ymax=7600,
        xtick={0.5,0.9,1.1},
        xticklabels={.5,.9,1.1},
        ytick={0,3584,7168},yticklabels={0,3.6K,7.2K},
        xlabel={Churn / capacity},ylabel={Prefix tokens},
        tick label style={font=\tiny},
        label style={font=\fontsize{4.5}{5}\selectfont},
        grid=major,grid style={gray!18},
      ]
        \addplot[very thick,green!55!black,mark=triangle*,mark size=1.6pt]
          table[x=churn,y=retire_retained] {figures/state_handoff_h100.dat};
        \addplot[very thick,blue!75!black,mark=*,mark size=1.5pt]
          table[x=churn,y=native_retained] {figures/state_handoff_h100.dat};
        \addplot[very thick,orange!90!black,mark=square*,mark size=1.5pt]
          table[x=churn,y=native_retained] {figures/state_handoff_gb200.dat};
      \end{axis}
    \end{tikzpicture}
    \par\vspace{-4pt}\tiny (b) State available
  \end{minipage}
  \caption{State handoff after intervening HBM churn.  Recovery is measured
  against tuned native abort with prefix caching; \retire retains the certified
  7168-token common prefix.  Points are medians of four paired quartets;
  whiskers span the four paired measurements.}
  \label{fig:kv-pressure}
\end{figure}
\endgroup

Figure~\ref{fig:kv-pressure} compares \retire with native abort plus vLLM prefix
caching on the same 8192-token predecessor and 7168-token common prefix.  Each
trial issues the switch, runs identical concurrent unique-prefix churn, and
then admits the successor.  H100 exposes two capacity steps: native caching
falls from 6560 to 2464 tokens at 0.85$\times$ churn and to zero at
0.95$\times$; recovery speedup rises to 4.82$\times$ and 6.87$\times$.
GB200 retains the full prefix through 0.95$\times$; at 1.00--1.10$\times$,
native caching loses it and \retire speeds recovery by 5.05--5.07$\times$.  Certified
retention keeps all 7168 tokens throughout.  A capacity-returning DRAM path
safely releases the source blocks and reaches up to 3.84$\times$/2.75$\times$
recovery on H100/GB200.

All 288 arm executions produce the same greedy successor tokens.  Every
\retire transition preserves the certified 448-block prefix, rejects retired
ownership, and reclaims only after writer closure.  The curve therefore
connects the authority abstraction to its conditional payoff: versioned
handoff matches a successful native cache and preserves fast recovery when
request-local cache ownership loses the successor's state.

\paragraph{No-revision cost.}
Selective guards
add 1.05\% median decode overhead (95\% bootstrap interval: 0.87--1.34\%) and
0.70\% for 4K prefill; the mixed eight-request test has no measurable slowdown.
A fresh interruption in the same campaign retains a 24.0$\times$ FENCE
reduction.  Selective placement also lowers the earlier always-armed decode
overhead from 5.25\% to 1.05\% and reduces guarded safe points from 35,190 to
1,080.  Appendix~\ref{app:details} reports the full matrix.

\begin{figure*}[t]
  \centering
  \includegraphics[width=0.92\textwidth]{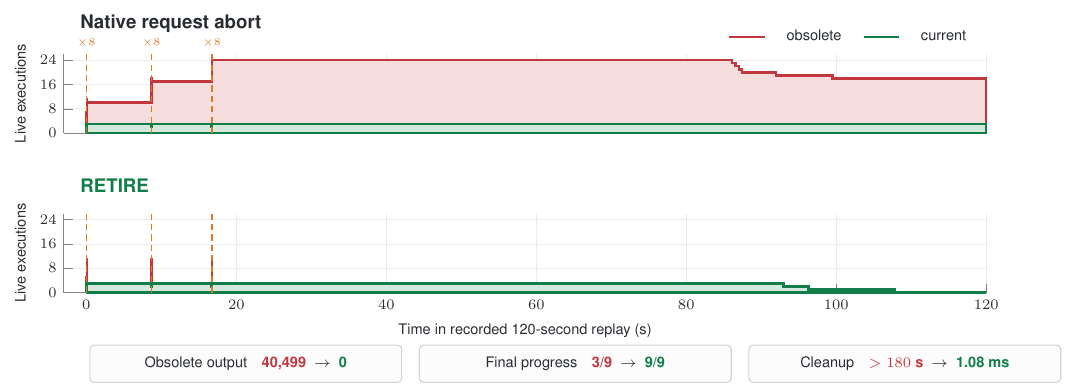}
  \caption{Recorded 120-second replay.  Identical axes show representative
  median trials.  Band thickness is current (green) or obsolete (red) live
  executions; the upper envelope is total live execution.  Orange lines mark
  three eight-revision bursts; summaries are medians of three trials.}
  \label{fig:recorded-trace}
\end{figure*}

\subsection{Repeated revisions, distributed execution, and isolation}
\label{sec:eval-real}

We now move from one handoff to a sequence of revisions, preserving the timing
and burst structure of recorded agent interruptions.  A fixed one-month
coding-agent trace contains 751 explicit
\texttt{turn\_aborted}$\rightarrow$\texttt{task\_started} transitions across
345 scopes.  We replay its densest 120 seconds: 24 transitions across three
concurrent scopes in three bursts, with inter-arrival coefficient of variation
3.57.  The serving state uses a controlled 8192-token prompt and zero prefix
overlap, isolating retirement behavior from content and cache reuse.

This replay asks whether a transition that works once can keep making progress
when another revision arrives before its predecessor has drained.  Such bursts
create chains of versions in which several intermediate executions may become
obsolete before emitting a token.  Current-version goodput, final-version
progress, obsolete publication, and cleanup latency jointly capture whether
the server follows the moving authority boundary.

Figure~\ref{fig:recorded-trace} makes the accumulating consequence of repeated
revision visible.  Native abort reaches 22 concurrently live obsolete
executions, delivers 40,499 obsolete tokens, leaves three streams at the
180-second cleanup limit, and advances only 3/9 final versions.  RETIRE confines
obsolete executions to millisecond-scale burst-local intervals, so they appear
as narrow spikes at this time scale; it emits no obsolete output, closes cleanup
in 1.08~ms, and advances all 9 final versions.  Current-version goodput is
54.0 versus 9.77 token/s, or 5.53$\times$, because service is redirected toward
the surviving versions.  Their median TTFT is 405~ms, compared with 87.0~s for
the three native final versions that produce output.  The full trace extraction
and policy matrix appear in Appendices~\ref{app:recorded-trace}
and~\ref{app:rate-matrix}.

\paragraph{Distributed scopes and tenant sharing.}
The remaining experiments widen the authority scope and add unrelated work.
We first repeat retirement and inheritance across tensor- and pipeline-parallel
participants, testing whether one transition closes every participating worker.

\begingroup
\setlength{\intextsep}{3pt}
\begin{table}[H]
  \caption{Repeated distributed closure.  Each row reports separate five-pair
  retirement and inheritance campaigns.}
  \label{tab:distributed}
  \centering
  \small
  \setlength{\tabcolsep}{3.5pt}
  \begin{tabular}{ccrrr}
    \toprule
    TP & PP & FENCE & Int.-to-succ. & Inheritance \\
    \midrule
    2 & 1 & 12.19$\times$ & 1.83$\times$ & 5.78$\times$ \\
    4 & 1 & 10.57$\times$ & 1.77$\times$ & 5.27$\times$ \\
    2 & 2 &  6.04$\times$ & 1.70$\times$ & 7.08$\times$ \\
    \bottomrule
  \end{tabular}
\end{table}
\endgroup

Every distributed retirement reconstructs the full
TP$\times$PP$\times$KV-group participant set, publishes no obsolete output,
and matches its clean oracle.  Every inheritance run adopts the same
4080-token certified physical prefix and matches its mode-matched reference.
Across the three topologies in Table~\ref{tab:distributed}, FENCE improves by
6.04--12.19$\times$, interruption-to-successor completion by
1.70--1.83$\times$, and inherited TTFT by 5.27--7.08$\times$.  In a TP=4
co-tenant run, superseding Tenant A leaves Tenant B's KV digest and block IDs
unchanged across all three seeds; B's median runtime changes by 1.0\% and its
cross-interrupt frame gap by 1.8\%.  Thus the same version boundary that closes
the controlled single-GPU lifecycle continues to hold under recorded bursts,
distributed participation, and shared serving resources.

\section{Implications for Revisable Serving}
\label{sec:discussion}

\paragraph{Invalidation and preservation are complementary.}
A revision changes which work is useful, not just which request should run
next.  Fast invalidation stops spending resources on the abandoned plan;
inheritance reduces the work required by its replacement.  Their shared
authority boundary lets the server optimize both together.  This is the central
consequence of treating revision as a serving workload.

\paragraph{Authority spans objects with different lifetimes.}
Output queues, issued computation, and reusable state do not finish together.
A common version decision gives them one logical boundary while allowing
physical cleanup to proceed independently.  Correctness depends on enforcing
that boundary wherever effects become visible, rather than on making every
object disappear at once.

\paragraph{Policies optimize within the invariant.}
Revision pressure changes the value of execution granularity and speculation;
prefix overlap and admission delay change the value of retention.  Schedulers
and memory policies can respond to these inputs while the same authority
checks govern publication and handoff.  Appendix~\ref{app:rate-matrix} explores
this separation through rate sweeps.

\paragraph{The runtime chooses; serving carries out the handoff.}
The runtime selects the successor context and handles external actions.
Serving determines when that successor can run and which state it can inherit.
This boundary lets agent behavior evolve while publication, isolation, and
efficient handoff remain systems responsibilities.

\section{Conclusion}
\label{sec:conclusion}

Agent revisions invalidate executions but leave useful work behind.
\retire redesigns serving accordingly: requests own resources,
versions own authority, and one protocol coordinates fast invalidation,
certified inheritance, and asynchronous reclamation.  Combining invalidation
and inheritance reduces paired revision-to-successor TTFT by a median 17.1\%.
Recorded-burst replay emits no obsolete output and advances 9/9 final versions;
distributed and co-tenant tests preserve the same authority boundary.
Versioned execution makes a change of plan a coordinated handoff: retire
obsolete work and preserve useful state.

\clearpage
\setlength{\textfloatsep}{\mlsysTextFloatSep}
\setlength{\dbltextfloatsep}{\mlsysDblTextFloatSep}
\bibliography{references}
\bibliographystyle{mlsys2025}

\clearpage
\appendix
\section{Methodology and Supplementary Evidence}
\label{app:details}

The appendix records the information needed to interpret and reproduce the
main results.

\subsection{Scope, Contract, and Integration}
\label{app:taxonomy}

The agent runtime resolves why an interruption occurred; serving receives one
version transition.  Table~\ref{tab:taxonomy-full} maps representative runtime
replacement decisions to that serving event.

\begin{table}[H]
  \caption{Interruption sources at the agent/serving boundary.}
  \label{tab:taxonomy-full}
  \centering
  \footnotesize
  \setlength{\tabcolsep}{3.5pt}
  \begin{tabularx}{\columnwidth}{@{}p{0.25\columnwidth}X@{}}
    \toprule
    \textbf{Source} & \textbf{Serving transition} \\
    \midrule
    User steering & Replace execution under revised goals or context \\
    \hdashline
    Tool or service & Invalidate work dependent on a failed or expired call \\
    \hdashline
    Environment & Replace work derived from a stale observation \\
    \hdashline
    \shortstack[l]{Validation and\\control} & Retire rejected output and dependent state \\
    \hdashline
    \shortstack[l]{Planning and\\concurrency} & Retire abandoned or losing branches as one scope \\
    \hdashline
    \shortstack[l]{Resource and\\operation} & Restart or migrate execution under a new policy \\
    \bottomrule
  \end{tabularx}
\end{table}

For scope $s$, the runtime supplies revised context $X'$ and advances a
monotonic epoch.  Every request, output, KV block, transfer, and worker
operation carries $(tenant,s,epoch,generation)$.  The checker requires: current
epoch at publication; writer closure and matching generation before
inheritance; complete participant agreement for distributed transitions; and
isolation of retired completions.  It records the authority linearization point
(ADVANCE), last old launch and writer receipts (FENCE), certified token and
block ranges (ALIGN), first successor publication (RESUME), and released pins
(RECLAIM).

In the vLLM implementation, EngineCore propagates the tag through scheduling,
worker metadata, attention, KV installation, and API output.  Long prefills
poll at bounded model boundaries; tensor-parallel participants retire
collectively.  ALIGN intersects the logical common prefix with writer
frontiers, fingerprints, generations, tenants, numerical mode, and KV-group
block boundaries.  The successor aliases certified physical blocks and
allocates its changed suffix under a fresh generation.  DRAM objects use the
same identity and digest checks.

\subsubsection{Coverage of prior systems}
\label{app:related-map}

Table~\ref{tab:related-protocol-map} locates prior mechanisms in the transition.
A filled circle is their primary locus; an open circle is a capability consumed
at that stage.

\begin{table*}[t]
  \caption{Existing mechanisms and version-transition stages.
  \(\bullet\): primary locus; \(\circ\): consumed capability.}
  \label{tab:related-protocol-map}
  \centering
  \small
  \setlength{\tabcolsep}{4.5pt}
  \begin{tabularx}{\textwidth}{@{}Xcccccc@{}}
    \toprule
    \textbf{Line of work} & \textbf{Upstream} & \textbf{Advance} &
    \textbf{Fence} & \textbf{Align} & \textbf{Resume} & \textbf{Reclaim} \\
    \midrule
    Behavioral interruption and stop enforcement
    \citep{wu2025reasoninginterruptible,rana2025agentchangebench,
    zou2026interruptbench,zhai2026revisable,khan2026stop} &
    $\bullet$ & $\circ$ & -- & -- & -- & -- \\
    \hdashline
    Agent and preemptive scheduling \citep{wu2026fastserve,luo2026agentix} &
    $\circ$ & -- & $\bullet$ & -- & $\bullet$ & $\circ$ \\
    \hdashline
    Intercept-aware serving \citep{abhyankar2024infercept} &
    $\circ$ & -- & $\bullet$ & $\circ$ & $\bullet$ & $\circ$ \\
    \hdashline
    KV paging and prefix sharing
    \citep{kwon2023pagedattention,zheng2024sglang} &
    -- & -- & -- & $\bullet$ & $\bullet$ & $\circ$ \\
    \hdashline
    Stateful caching and tiered restoration
    \citep{gao2024cachedattention,yu2025pensieve,gao2025hcache,
    qin2025mooncake,ma2026leyline} &
    -- & -- & $\circ$ & $\bullet$ & $\bullet$ & $\bullet$ \\
    \hdashline
    Rollback-consistent KV retention \citep{zhang2026aborted} &
    $\circ$ & $\circ$ & $\bullet$ & $\bullet$ & $\circ$ & -- \\
    \midrule
    \textbf{\retire versioned execution} &
    $\circ$ & $\bullet$ & $\bullet$ & $\bullet$ & $\bullet$ & $\bullet$ \\
    \bottomrule
  \end{tabularx}
\end{table*}

\subsection{Workloads and Reproducibility}

Single-GPU experiments use H100 PCIe and GB200 systems with CUDA 13.1.  Unless
stated otherwise, results are paired trials summarized by medians; speedups
are computed before display rounding.  Correctness trials compare the authoritative successor
with a clean execution under the declared numerical mode.  Timing begins at
the version-switch signal and ends at the named successor event; identical
churn or arrival schedules are replayed across arms.

\subsubsection{Recorded agent trace}
\label{app:recorded-trace}

The extractor identifies a transition when \texttt{turn\_aborted} is followed
by \texttt{task\_started} in the same thread within one hour.  A fixed one-month
trace contains 751 transitions across 345 scopes.  Its densest 120-second
window has 24 transitions across three scopes and three eight-transition
bursts (inter-arrival coefficient of variation 3.568).  Replay preserves these
times and uses an 8192-token controlled state with zero prefix overlap, so this
experiment isolates repeated retirement from cache reuse.

\begin{table}[H]
  \caption{Recorded-trace replay; medians of three trials.}
  \label{tab:supp-recorded}
  \centering
  \small
  \setlength{\tabcolsep}{3pt}
  \begin{tabularx}{\columnwidth}{@{}Xrrr@{}}
    \toprule
    \textbf{Path} & \textbf{Quantum} & \textbf{Goodput} & \textbf{Obsolete} \\
    & \textbf{(tokens)} & \textbf{(token/s)} & \textbf{tokens} \\
    \midrule
    Native abort & 1024 & 9.77 & 40,499 \\
    Native abort & 8192 & 8.18 & 40,123 \\
    Always armed & 4096 & 54.6 & 0 \\
    Selective guards & 1024 & 54.5 & 0 \\
    Selective guards & 8192 & 54.0 & 0 \\
    Rate-only selector & 8192 & 53.9 & 0 \\
    \bottomrule
  \end{tabularx}
\end{table}

Both native configurations hit the 180-second cleanup limit and allow 3/9
final versions to publish.  Every versioned configuration completes cleanup,
advances 9/9 final versions, and publishes zero obsolete tokens.

\subsubsection{Revision-rate matrix}
\label{app:rate-matrix}

The supplementary sweep crosses revision rates 0, 0.05, and 0.2 per scope per
second with concurrencies 4 and 8, five execution policies, and three
repetitions per cell (108 audited points).  The current-goodput-optimal native
quantum is 8192 tokens at zero and moderate revision pressure, but moves to
1024 at rate 0.2 for both concurrencies.  The rate-only selector follows this
direction, characterizing how the preferred quantum changes with revision
pressure.

\subsection{Correctness Closure}

Integrated long-prefill, draft--verify prefill, active speculative-decode, and
heterogeneous-KV-group trials publish zero predecessor output and match their
clean successors.  ALIGN crosses five prompt seeds with LCPs of 4080 and
4095 tokens, yielding ten seed--LCP configurations.  Each configuration has
three arms---cold, clean-prefix, and completed-predecessor---with two
repetitions per arm: 60 successor measurements per execution mode, or 120
across ordinary and batch-invariant modes.  Clean-prefix and
completed-predecessor sources produce identical outputs in all ten
configurations in both modes.  Batch-invariant execution additionally makes
cold and reuse chosen-token log probabilities identical.

A positive page-fault control retains block IDs and cache counts while zeroing
adopted GPU pages.  It changes output in 3/3 trials; restoring the certified
bytes recovers exact output in 3/3.  This separates physical-state dependence
from batch-dependent floating-point variation.

Distributed retirement reconstructs the complete TP$\times$PP$\times$KV-group
participant set.  All 15 retirement trials publish zero obsolete output and
match clean oracles; all 15 inheritance trials adopt the same 4080-token
physical prefix and match mode-matched references.  In three TP=4 co-tenant
seeds, superseding Tenant A leaves Tenant B's KV digest and block IDs unchanged.
Tenant B's median runtime and cross-interrupt frame gap change by 1.0\% and
1.8\%, respectively.

\paragraph{Bounded retention and quota fallback.}
Pins are accounted for by tenant and released after successor adoption is
verified; successor references then keep the shared pages live.  Both
three-seed isolation campaigns also exercise a zero-block pin quota.  Every
case selects cold recovery with zero inherited tokens, matches its clean
successor, publishes no obsolete output, and safely reclaims the predecessor.

\subsection{Supplementary Performance Results}
\label{tab:supp-retire}

Table~\ref{tab:retire-paths} reports the complete integrated retirement matrix.
The H100 runs use an 8192-token prompt, a 4096-token active unit, and a 5-ms
injection delay.  Qwen3-32B uses BF16; the speculative arm enables three n-gram
draft tokens, and GPT-OSS-20B supplies sliding and full-attention KV groups.

\begin{table}[H]
  \caption{Integrated retirement results.}
  \label{tab:retire-paths}
  \centering
  \small
  \setlength{\tabcolsep}{3pt}
  \begin{tabularx}{\columnwidth}{@{}Xrrr@{}}
    \toprule
    \textbf{Path} & \textbf{Pairs} & \textbf{FENCE} & \textbf{Int.-to-succ.} \\
    \midrule
    Long prefill & 3 & 24.9$\times$ & 1.48$\times$ \\
    Draft--verify prefill & 3 & 51.3$\times$ & 1.42$\times$ \\
    Active spec. decode & 5 & 9.99$\times$ & 1.12$\times$ \\
    Two KV groups & 3 & 7.01$\times$ & 1.22$\times$ \\
    \bottomrule
  \end{tabularx}
\end{table}

The no-revision campaign uses six loaded burn-in groups, two warmups, and ten
paired groups per arm.  Selective arming reduces the earlier always-armed
decode overhead from 5.25\% to 1.05\% and safe points from 35,190 to 1,080.

\begin{table}[H]
  \caption{Selective-path no-revision cost; paired 95\% bootstrap intervals.}
  \label{tab:supp-steady}
  \centering
  \small
  \setlength{\tabcolsep}{3pt}
  \begin{tabularx}{\columnwidth}{@{}Xrr@{}}
    \toprule
    \textbf{Workload} & \textbf{Change} & \textbf{95\% interval} \\
    \midrule
    1K prefill, concurrency 1 & +0.140\% & [$-$3.27, +4.76]\% \\
    4K prefill, concurrency 1 & +0.697\% & [+0.198, +1.46]\% \\
    128-token decode, concurrency 1 & +1.05\% & [+0.873, +1.34]\% \\
    Mixed, concurrency 8 & $-$0.646\% & [$-$1.74, +0.381]\% \\
    \bottomrule
  \end{tabularx}
\end{table}

\paragraph{KV recovery and tiering.}
\label{app:kv-pressure}
The strong-baseline control uses a fully computed 4096-token shared prefix,
zero pressure, immediate successor submission, and tuned native cancellation
with prefix caching.  Five trials at each of three interruption offsets produce
45 accepted executions per GPU.  Native caching and versioned retention both
reuse exactly 4096 tokens; versioned execution adds 1--2\% latency across H100
PCIe and GB200.

The pressure campaign uses an 8192-token predecessor, a block-aligned
7168-token common prefix, and four paired trials at each pressure point.  Each
quartet runs native cold restart, native abort with prefix caching, versioned
HBM retention, and capacity-returning DRAM offload under identical concurrent
churn.  Table~\ref{tab:handoff-policy} reports both successor recovery and the
full interruption interval.  Recovery begins when the churn phase completes;
the full interval begins at the version switch.  Speedups use the corresponding
native-prefix execution in the same trial.

\begin{table*}[t]
  \caption{State-handoff policy under HBM churn; medians of four paired
  quartets.  Recovery and full-interval speedups are relative to native abort
  with prefix caching.}
  \label{tab:handoff-policy}
  \centering
  \small
  \setlength{\tabcolsep}{5pt}
  \begin{tabular}{lrrrrrr}
    \toprule
    GPU & Churn & Native prefix & \multicolumn{2}{c}{Recovery speedup} &
    \multicolumn{2}{c}{Full-interval speedup} \\
    \cmidrule(lr){4-5}\cmidrule(l){6-7}
    & & retained & HBM & DRAM & HBM & DRAM \\
    \midrule
    H100 PCIe & 0.50$\times$ & 7168 & 0.998$\times$ & 0.798$\times$ & 1.001$\times$ & 0.948$\times$ \\
    H100 PCIe & 0.75$\times$ & 6560 & 1.58$\times$ & 0.786$\times$ & 1.015$\times$ & 0.927$\times$ \\
    H100 PCIe & 0.80$\times$ & 6560 & 1.59$\times$ & 0.699$\times$ & 1.021$\times$ & 0.942$\times$ \\
    H100 PCIe & 0.85$\times$ & 2464 & 4.82$\times$ & 2.29$\times$ & 1.139$\times$ & 1.061$\times$ \\
    H100 PCIe & 0.90$\times$ & 2464 & 4.93$\times$ & 2.31$\times$ & 1.147$\times$ & 1.067$\times$ \\
    H100 PCIe & 0.95$\times$ & 0 & 6.87$\times$ & 3.27$\times$ & 1.190$\times$ & 1.118$\times$ \\
    H100 PCIe & 1.00$\times$ & 0 & 6.51$\times$ & 3.84$\times$ & 1.176$\times$ & 1.117$\times$ \\
    H100 PCIe & 1.05$\times$ & 0 & 6.79$\times$ & 3.25$\times$ & 1.189$\times$ & 1.121$\times$ \\
    H100 PCIe & 1.10$\times$ & 0 & 6.79$\times$ & 3.23$\times$ & 1.172$\times$ & 1.113$\times$ \\
    \midrule
    GB200 & 0.50$\times$ & 7168 & 0.993$\times$ & 0.500$\times$ & 0.999$\times$ & 0.985$\times$ \\
    GB200 & 0.75$\times$ & 7168 & 0.986$\times$ & 0.532$\times$ & 1.000$\times$ & 0.990$\times$ \\
    GB200 & 0.80$\times$ & 7168 & 0.989$\times$ & 0.521$\times$ & 0.996$\times$ & 0.989$\times$ \\
    GB200 & 0.85$\times$ & 7168 & 0.991$\times$ & 0.500$\times$ & 1.000$\times$ & 0.988$\times$ \\
    GB200 & 0.90$\times$ & 7168 & 0.990$\times$ & 0.518$\times$ & 1.000$\times$ & 0.991$\times$ \\
    GB200 & 0.95$\times$ & 7168 & 0.988$\times$ & 0.517$\times$ & 1.000$\times$ & 0.994$\times$ \\
    GB200 & 1.00$\times$ & 0 & 5.06$\times$ & 2.65$\times$ & 1.019$\times$ & 1.011$\times$ \\
    GB200 & 1.05$\times$ & 0 & 5.07$\times$ & 2.75$\times$ & 1.018$\times$ & 1.012$\times$ \\
    GB200 & 1.10$\times$ & 0 & 5.05$\times$ & 2.67$\times$ & 1.017$\times$ & 1.011$\times$ \\
    \bottomrule
  \end{tabular}
\end{table*}

All 288 executions produce identical greedy successor tokens.  HBM retention
and DRAM offload recover the complete 7168-token prefix in every trial.  The
offload path returns all 512 predecessor blocks to the allocator; churn
overwrites their former slots, after which restoration allocates and
conditionally publishes the 448 certified blocks under the successor's
generation.  Median D2H/H2D times span 221--273/270--308 ms on H100 PCIe and
95.7--117/126--147 ms on GB200.  Integrity, generation, ownership, and physical
remapping checks pass in every trial.

\paragraph{Distributed timing.}
The three five-pair TP$\times$PP campaigns use a 16,384-token retirement prompt
and a 4080-token aligned inherited prefix.  For 2$\times$1, 4$\times$1, and
2$\times$2, baseline/RETIRE FENCE medians are 2638.7/216.5,
1381.2/130.6, and 2680.0/443.6 ms.  Corresponding pinned/cold inherited TTFTs
are 44.8/258.7, 27.9/146.9, and 35.3/249.7 ms.

The accompanying artifact contains the implementation, trace and workload
generators, paired-run harnesses, contract checker, normalized summaries, and
machine-readable measurements used in the paper.

\end{document}